# Control Flow Change in Assembly as a Classifier in Malware Analysis

Andree Linke
School of Computer Science
University College Dublin
Ireland
andree.linkee@ucdconnect.ie

Nhien-An Le-Khac
School of Computer Science
University College Dublin
Ireland
an.lekhac@ucd.ie

*Abstract*—As currently classical malware detection methods based on signatures fail to detect new malware, they are not always efficient with new obfuscation techniques. Besides, new malware is easily created and old malware can be recoded to produce new one. Therefore, classical Antivirus becomes consistently less effective in dealing with those new threats. Also malware gets hand tailored to bypass network security and Antivirus. But as analysts do not have enough time to dissect suspected malware by hand, automated approaches have been developed. To cope with the mass of new malware, statistical and machine learning methods proved to be a good approach classifying programs, especially when using multiple approaches together to provide a likelihood of software being malicious. In normal approach, some steps have been taken, mostly by analyzing the opcodes or mnemonics of disassembly and their distribution. In this paper, we focus on the control flow change (CFC) itself and finding out if it is significant to detect malware. In the scope of this work, only relative control flow changes are contemplated, as these are easier to extract from the first chosen disassembler library and are within a range of 256 addresses. These features are analyzed as a raw feature, as n-grams of length 2, 4 and 6 and the even more abstract feature of the occurrences of the n-grams is used. Statistical methods were used as well as the Naïve-Bayes algorithm to find out if there is significant data in CFC. We also test our approach with real-world datasets.

## I. Introduction

The world of computer crime is constantly expanding. Due to constantly new tech-nology is invading our lives, the opportunities of making money by exploiting tech-nologies' vulnerabilities rise in the same way. At the same time, classical antivirus (AV) products seem to fail against new coded malware [1], which incorporates rootkit technologies and gets encoded to subvert AV products. Classical AV relies greatly on file signatures, providing which is a reactive process of finding a malware, creating a signature (for example by hashing or extracting byte sequences) and pushing these signatures into file/system scanners. For institutions like the police or military, this approach is no more feasible, as the attackers have become more proficient and equipped and institutions face a constant stream of sophisticated attacks.

Therefore, new automated methods of discern between wanted software (so-called "goodware") and unwanted software ("malware") ought to be explored to battle the stream of malware. Interesting approaches have been taken in the past and lead to systems for automatic detection and categorization of malware, such as sandboxes or intrusion prevention systems. Current approaches have been taken to use statistical analysis [2] or machine learning [3] to find discriminators for categorization. As the analysis of microprocessor operation code (opcode) has been subject of some research and some approaches have been suggested for analysing the control flow, in this paper we focus on relative change of control flow in static disassembly. This approach has not been proposed in the literature yet, so our work aims on testing if the use of control flow change can be used to differentiate between goodware and malware. The precondition for our approach is that the software in question is not packed, encrypted or encoded. Software unpacking, decryption or decoding is beyond the scope of this work, however, simple steps in sorting out such samples have been taken.

The rest of this paper is organised as follows: Section 2 shows background of our research and related work in this area. We present our approach in Section 3. We describe and analyse results in Section 4. Finally, we conclude and discuss on future work in Section 5.

## II. Background

### A. Windows PE files

The PE file format is the main format of Microsoft Windows executable files, dynamic link libraries and object code. It contains all information needed for the program loader of the Windows operating system to build the process object, the memory layout and needed library call structures. It is derived from the Unix COFF file format. The supported architectures of the PE file format are IA-32, IA-64, x86-64 and ARM. This work focuses on the IA-32 architecture. The full documentation of the PE file format can be found in Microsofts "Microsoft PE and COFF Specification" [4]. The code of the executable can be extracted from the sections part of the PE file in raw form using the section table information. In this paper, the executable segments of a program are extracted from the PE file using the "pefile"-library for python by Ero Carrera [5].

*B. Interactive DisAssembler (IDA)*

Disassembling a compiled program is the process of translating an executable pro-gram into an equivalent mnemonic representation, which is human-readable. It is the inverse operation of an assembler. Generally, disassembling is done by having a reference of the opcode bytes as the "Intel® 64 and IA-32 Architectures Software Developer Manuals" (Intel Corporation 2014) and by looking up the opcodes, assigning the appropriate mnemonic with the corresponding operands. In this work, the process of disassembling is done by the python library "distorm3" [6]. The disassembler library is used to distinguish the operations against each other, as x86 operations are not aligned to a specific length and therefore not trivially detectable. The second disassembler used is the Interactive DisAssembler (IDA) Pro by Ilfak Guilfanov[7] in Version 6.2.0111006. IDA supports multiple architectures and binary formats. The disassembler was modified by an .idc script to automatically provide an opcode listing as well as the mnemonic representation. Figure 1 shows an excerpt of the listing exported by IDA.

```
.text:01001316                 var_4           = dword ptr -4
.text:01001316                 hInstance       = dword ptr  8
.text:01001316                 hPrevInstance   = dword ptr  0Ch
.text:01001316                 lpCmdLine       = dword ptr  10h
.text:01001316                 nShowCmd        = dword ptr  14h
.text:01001316
.text:01001316 8B FF                           mov     edi, edi
.text:01001318 55                              push    ebp
.text:01001319 8B EC                           mov     ebp, esp
.text:0100131B 81 EC 1C 02 00 00               sub     esp, 21Ch
.text:01001321 A1 00 30 00 01                  mov     eax, ___security_cookie
.text:01001326 33 C5                           xor     eax, ebp
.text:01001328 89 45 FC                        mov     [ebp+var_4], eax
.text:0100132B 8B 45 08                        mov     eax, [ebp+hInstance]
.text:0100132E 53                              push    ebx
.text:0100132F 56                              push    esi
.text:01001330 8B 75 10                        mov     esi, [ebp+lpCmdLine]
.text:01001333 57                              push    edi
.text:01001334 33 DB                           xor     ebx, ebx
```

Fig. 1. IDA listing, excerpt from *AdapterTroubleshooter.exe*

*C. Control flow change (CFC) and n-gram*

In our approach, it is assumed that control flow changes differ between malware and goodware. As malware has to test many variables in an infected system, such as testing for antivirus software, potentially interesting data or evaluating data collected for example by keyboard sniffers, in this work it is expected that CFC patterns differ between goodware and malware. If this is the case, statistical methods will show if this difference can be used to classify unknown software as goodware and malware. The methods used in our approach to find a discriminator between malware and goodware include the statistical values of the median, the variance, the variance coefficient and the spread.

Another approach to find a discriminator is to use a classifier. The Naive Bayes Classifier[8] is expected to perform well, being relatively simple to implement and having good detection rates. The Naive Bayes Classifier is derived from the Bayes' Theorem. In this work, the CFC features are treated like words as input for the Naive Bayes Classifier. The classifier is trained with goodware and malware and afterwards, selected goodware and malware samples are tested using the classificatory.

Also the n-gram method is used in this work. An n-gram is a contiguous sequence of letters, a substring of a larger word or text. The "n" is designating the length of the contiguous sequence. The word "word", for example, can be broken into the 2-grams (or bigrams) "wo", "or" and "rd".

*D. Related work*

Bilar [2] describes the method of gathering statistical data about opcode distribution in assembly, using this to predict if a program is malware. The opcode frequencies of 67 malware samples and 20 non-malicious programs were evaluated. Therefore the malware was classified by unknown methods into the class kernel-mode rootkit, usermode rootkit, tool, bot, trojan, virus and worm. The instructions were counted by the IDAPro disassembler plugin InstructionCounter and statistically examined by a Java program. The analysis was performed inside a virtual machine using VMWare Player to contain the malware samples. The common opcodes (opcodes frequently seen in software) did not prove to be a strong predictor for malware, "about 70% of the cells exhibited similar, 30% higher and 10% lower opcode frequencies", where "cells" are the malware opcode frequencies in the different malware "groups" like trojan or rootkit. The results also indicate a significant difference between the mal-ware classes and goodware in the more infrequent opcodes, classifying them as a strong predictor for malware. As this approach has proven the value of statistical data in opcode distribution, this work incorporates the idea.

This method is refined by Santos et al. [3], who use machine learning methods to detect unknown malware samples. It is proven that machine learning can successfully be applied to opcode frequency data to distinguish between malware and good software with low false positive ratios. The frequency of opcode sequences is used as a vector representation of the program executables. This approach is used in this work as well with the frequency of the extracted n-grams. Methods used are the Decision Tree classifier, different types of Support Vector Machines, K-Nearest Neighbours with K varying from 1 to 10, the Naive Bayes Algorithm and Bayesian Networks. The opcode sequence contemplated is 1 or 2, as well as the combined likelihoods of both. It is proven that the method used "provides a good detection ratio of unknown malware while keeping a low false positive ratio" [2].

A similar approach is taken by Kang et al. [9]. This approach focuses at the mnemonic rather than the opcode and therefore subsumes similar operations (for example xor 0x30-0x35) presenting stronger statistical data due to abstraction. Kang et al. make use of Intels PIN library[10] for the dynamic approach, so the assembly is extracted out of the running program. The library runs executable code and breaks on every branch, so custom code can be run. This is an approach to counter packed malware.

Ding et al. [11] further improve this method by looking at the control flow changes. Code blocks ("basic blocks") can be identified and the exact opcode sequence and therefore distinct execution paths can be recovered. These execution paths can be split into n-grams of consecutive opcodes using the sliding window method. A database of these n-grams can be built, optimizing the n-gram size (although only 3-grams are covered for performance reasons). The n-grams with the highest information gain are selected as features and used to categorize malware.

Another method of malware detection is by using n-gram signatures of files [12]. N-grams, which are a substring of length n of a string, are extracted out of files and used as an input for a k-nearest-neighbour algorithm, which then classifies a test set for prior learned sample sets of goodware and malware. Unfortunately, the exact feature extracted is not clear. The extracted n-gram data provided to be a usable classification feature with a best ratio without false positives of 74,37% malware detection ratio using 4-grams and 17 most alike malware files (nearest neighbours). Therefore, in our approach, we focus on 2-grams, 4-grams and 6-grams.

III. TOWARD A NEW APPROACH OF CLASSIFY MALWARE

A. Problem statement

In Section 2, the current approach in literature adopted has been based on the opcode frequency, mnemonic frequency and opcode n-grams to classify programs as malware and goodware. By extracting statistical data about both groups, it is shown that there is a difference in the statistical distribution of certain features between malware and goodware. It is also shown that all approaches, the classical statistical one [2] and the advanced ones, for example Naive Bayes or clustering [3][12], are promising. However, there are more features in assembly code which can be extracted. Our approach focuses on the change in control flow, namely the relative jump and call opcodes and their parameters. These features are extracted from samples of malware and goodware and it is tested if relative control flow change can be used as a discriminator between these groups. If it is, it will be shown how statistics about control flow change perform against the statistical analysis of opcodes. This exact feature has not been subject to research before, therefore multiple methods are tested. The objective of our research is to test if there is a significant difference in control flow change between goodware and malware. Therefore it is assumed that CFC patterns differ between goodware and malware. If this is the case, statistical methods will show if this difference can be used to classify unknown software as goodware and malware. The feature is furthermore thought of as a classificator for program families (both goodware and malware) or library use in programs, which will be subject of further research. Hereby, it is known that the expression of control flow change lies mostly in the hands of the compiler, but as malware authors tend to stick with the compiler they used before, this is seen as a minor problem for this work but a topic for further research if successful.

Also the use of anti-debugging techniques in malware ought to pose a problem. This problem is to be addressed by improving disassembling techniques. In prior re-search this has been a problem too. For the disassembly process, IDA was used in most of the works where the disassembler was mentioned. As the opcodes were extracted as features and a significant result has been found, IDA is also expected for disassembly to be reliable enough to extract statistical data and to pose a problem to be addressed in disassembly research. These challenges are connected to another challenge in disassembly, the occurrence of data within the disassemble area. As this occurred during the research the disassembler was changed from distorm3 to IDA. Furthermore, the control flow changes are used as base text for the n-gram method. N-grams have proven to be a usable feature by Santos et al. [12], therefore this approach has been chosen to abstract from the raw CFC data for testing the n-gram method further. The extracted n-grams are reviewed using methods as in the former research.

B. Proposed Approach

The Zeus and Citadel malware among other less known samples for this work have been obtained from the malware archive of the University of Bonn. Also the parts of the malware database of contagio.blogspot.de [15] and nothink.org [16] are also chosen in our analysis. According to the archive owners, the malware chosen from their archives ought to be unpacked. All files chosen for this work are in the Microsoft PE format.

As the sample set for goodware, we chose executable smaller than 1MB and not named "x86_microsoft*" of the windows system directory of a Windows 7 SP 1 x32 with a patch level of 20.02.2014. These executables represent programs changing parameters in the operating system and accessing operating system functions and are therefore thought to be most similar to malware in wanted, "good" software. So if significant CFC differences can be found, the CFC approach should work even better with software not performing work on operating system level. The files "named x86_microsoft*" were omitted because these files are remains of Microsoft Windows updates and therefore not representative for normal windows applications.

At first every sample was tested if it is a PE/COFF executable by the pefile library. Nonparseable executables were discarded/skipped. The section characteristics in the PE header were tested for executable sections and these sections were extracted. The extracted sections were then tested for entropy [11]. According to "Using Entropy Analysis to Find Encrypted and Packed Malware" [13], "Using a 99.99 percent confidence level, executables with an average entropy and a highest entropy block value of greater than 6.677 and 7.199, respectively, are statistically likely to be packed or encrypted" [13]. Therefore, executable sections with entropy larger than 6.677 were also discarded automatically. Although this approach ignores the possibility of false negatives, it is accepted due to the time consumption of manual analysis and the error handling of IDA. False positives in the former methods just would lead to a piece of code not being taken into account for good/malware. This also is acceptable, because if the base of data is too small, more samples would be used. The code sections left over by these filters are automatically disassembled using distorm DecodeGenerator in Version 0x030300. The code is returned as a list of offset, size, instruction in human readable form and hexdump, so it can be used to match opcodes in hexadecimal as well as mnemonics. The single operations are then analysed by opcode. If an opcode causes a control flow change the address is extracted and appended with its control flow change length to a list for this opcode. The development and analysis took place on a virtual machine using VMWare Workstation 10.0 on a Microsoft Windows 7 x64 host. The VM runs elementary OS Luna kernel 3.2.0-63. Used python version is 2.7.3, but the programs were also tested on 2.7.5.

As problems were encountered during the work, a second disassembler was chosen to cope with strings and data in the executable segment. Many zero-length control flow changes and therefore many zero n-grams were seen. As zero-length jumps are seen in normal code too (but very scarcely, possibly for hot-patching reasons), the number of these CFC observed was very high. Some samples were loaded manually into IDA Pro and searched for the zero-length CFC in question. As none have been seen, it was observed that two-byte unicode strings featured some of the CFC in question, such as "0x75 0x00" (jump on not equal 0) are part of for example "0x00 0x75 0x00 0x70", forming unicode "up". To cope with this disassembly misinterpretation, another disassembler was used and some of the feature extraction and statistics code were rewritten. The product chosen is the IDA Pro Version 6.2.0111006 Disassembler by hex-rays. IDA was run on Windows 7 x64 SP1 Updates to 26.02.2015. For feeding the goodware and malware samples into IDA, python 2.7.5 for Windows was used. The listings were collected by the script and then copied to the Linux analysis VM. As IDA by itself does not support export of assembly listings including opcodes as parsable text, the following changes were made.

For our work, only relative control flow changes are contemplated. The structure generated by the collector program is a dictionary of opcode bytes as key, containing a list containing the opcode frequency and the address and jump length pairs. In case of distorm3 this structure is wrapped in a dictionary with the section- and filename as keys, containing also the relative virtual address (RVA), lines of code and entropy for further analysis.

Using IDA, the structure is the same; however neither RVA nor the entropy is calculated. Instead, RVA is set to 0 and entropy is set to 1 so it is possible to find out what disassembler was used by reading the output file. The whole dictionary is saved to a binary file in pickle format, so it can be retrieved to the original python data structures by other programs. The pickle module is an algorithm to serialize python objects, in this work it is used to write and read python objects to and from files. In a separate program, statistical features are calculated for the instructions "jump on condition" and relative call (0xe8) using this data structures. For each executable section, the following statistical numbers are calculated [8][14]: spread (smallest to largest value), the variance, the medians, the median divided by the maximum of the values minus the minimum of the values, the variance coefficient. All statistical data is sorted by value. To check if there is a significant difference in the distribution of the statistical values of goodware and malware, the Spearmans rank correlation coefficient (Spearman's Rho) test is applied to the distinct value pairs (positive and negative). Spearman's Rho was chosen because the data is in form of a continuous distribution. If the distribution of values of malware is a monotonous function of the distribution of values of goodware the value does not qualify as a statistically significant discriminator between the groups of malware and goodware. Therefore, the zero hypothesis was chosen as a perfect correlation between malware and goodware of all statistical values.

The second approach for finding a discriminator was to use a Naive Bayes Classifier to test unknown software against the sample datasets. An own Naive Bayes Classifier was developed on base of an implementation of Thomas Uhrig [19], which then was trained with the sample sets (574 goodware and 94 malware samples) of raw jump length data. Test software (both malware and goodware) was then tested against the data. The input for the Naive Bayes Algorithm was the length of the data of the raw control flow change of jump on condition (jcc, e.g. jz, jump on zero). The sample set of malware and goodware samples has been used to train the Naive Bayes Algorithm (training set) and 14 elements of malware and 4 elements of goodware programs have been tested against the learned data. The later chosen approach is based on the extraction of n-grams of words from a text, correspondingly jump length sequences of a code segment 25. The n-gram is the continuous sequence of n jump lengths from the list of control flow changes. The n-grams are extracted by choice of the user and written to a Sqlite database for easy comparison and further statistical handling. Two sample databases were chosen, one containing n-grams of 626 samples of goodware and another one containing n-grams of 95 samples of malware. Test samples were treated using the same process and were put into a separate database. The categorization test searches for occurrence of the to-test n-grams in the sample datasets and showed if conformities are found.

A second program extracts the n-grams by file; saving it in a pickle format file so further processing can be done with the n-grams belonging to the single files. The diversity in data formats (Sqlite versus pickle) is due to Sqlite being processable by non-python software for further research. For the further tests, the malware training dataset was increased to 535 samples, including samples of the then-new Equation campaign. Formerly done tests were repeated but the results did not differ significantly from the former results. Both database formats were used with the Naive Bayes Classifier, modified to classify n-grams. The classifier was trained with the goodware and malware Sqlite databases, drawing likelihood data for the whole set of n-grams. Then the test sample was tested against the classifier. Therefore, the n-grams of the single files needed to be identifiable. The classification likelihood for the goodware and malware classes is shown as in the first approach to the Bayesian Classifier. For this experiment, 2-, 4-, and 6-gram sets were created and tested. It was also tested how many n-grams are exclusive to goodware or malware and how many occur in both sample sets. The approach of using the n-grams themselves as a feature was omitted due to the unsatisfying results.

The further approach was to count the occurrences of the n-grams, choose a varia-ble length of the most frequently occurrences and use the frequency as a feature to feed into Naive Bayes for classification. The occurrences of the n-grams themselves also were counted for use as an even more indirect feature.

## IV. EXPERIMENTS AND ANALYSIS OF RESULTS

In this section, we describe our experiments of classification goodware and malware based on two methods: statistical and Naïve Bayes analysis on the length of jump on condition (jcc)..

### A. Statistical analysis

For the statistical analysis, the following measures of jcc have been chosen: the spread, the variance, the medians, the median divided by the maximum of the values minus the minimum of the values, the variance coefficient and the frequencies. As an overview, the average values of these statistical numbers are presented in Table 1.

TABLE I. STATISTICAL ANALYSIS OF JCC FOR GOODWARE AND MALWARE

|  | *Goodware* | *Malware* |
|---|---|---|
| *Spread* | 116 | 124 |
| *Scatter* | 27.02 | 31.63 |
| *Medians* | 18.18 | 20.99 |
| *Medians/Spread* | 0.17 | 0.17 |
| *Variance Coefficient* | 0.94 | 0.97 |
| *Frequencies* | 974 | 3373 |

All of these values were collected from the selected goodware samples and com-pared against the values collected from the malware samples. We found high correla-tion between goodware and malware for these statistical measures. In Table 1, the difference of all measures between the goodware and the malware is relative small except the frequencies. So, at the first stage, opcode frequency can be used as a classifier of goodware/malware, but this has been covered in prior research.

### B. Naïve-Bayes

For the Naive Bayes Classifier 14 samples of malware test data and 4 pieces of goodware test data were chosen. The classifier is trained for the dataset of jcc instruction. The training set included 620 samples of goodware and 94 samples of malware. The results of the jcc tests of the Naive Bayes Classifier are shown in Figure 2 and 3.

| filename | prob. good | prob. bad | length of data |
|---|---|---|---|
| aspnet_compiler.lst | 1.85E-028 | 4.97E-016 | 140 |
| bootcfg.lst | 6.08E-056 | 5.89E-034 | 1595 |
| AdapterTroubleshooter.lst | 8.80E-006 | 2.25E-006 | 110 |
| BrmfRsmg.lst | 8.77E-032 | 1.88E-019 | 641 |

Fig. 2. Naïve Bayes classifier likelihoods for jcc goodware

| filename | prob. good | prob. bad | length of data |
|---|---|---|---|
| bfb27f14234725a8f0146957953205ed.lst | 2.25E-078 | 7.12E-045 | 4659 |
| 1fd05f3185733f03e71543c0e27d7740.lst | 2.78E-077 | 3.37E-044 | 4807 |
| 0ec0f4be802b39a51c69bb0307a9629e.lst | 3.17E-077 | 3.64E-044 | 4863 |
| ff230a338ac820d73770411bab0013df.lst | 1.16E-074 | 1.84E-042 | 4393 |
| 6a8d6aec6af71a9ef65f4e1ac44da94b.lst | 5.84E-078 | 1.47E-045 | 4804 |
| 8f316e19714bad573af0f1116115cc33.lst | 2.78E-077 | 3.37E-044 | 4805 |
| 7e941465c1b5396697e9a2bebefe775c.lst | 8.15E-078 | 9.71E-045 | 4915 |
| 31f192e2e086723408ffc013bf546cbd.lst | 3.05E-078 | 6.58E-045 | 4815 |
| 9ecc6d7904710fd0b45926ae535a2529.lst | 2.78E-077 | 3.37E-044 | 4806 |
| 6fcc3a8b55376793f2985efbcb0123c8.lst | 4.68E-078 | 3.97E-044 | 4861 |
| 4c6a9aaaae5ec8cbb430a969bb17849c.lst | 8.81E-077 | 7.62E-044 | 4533 |
| 2fa2cbb2d273ab21aa1e10a6b314484f.lst | 2.78E-077 | 3.37E-044 | 4802 |
| 5a304d1f64643b9501f5d43a67460ca5.lst | 2.78E-077 | 3.37E-044 | 4805 |
| 425554e39f37bb5af1d8280e6fdd563d.lst | 2.20E-077 | 4.58E-044 | 4588 |

Fig. 3. Naïve Bayes classifier likelihoods for jcc malware

As the n-grams were extracted from the raw CFC data, the next step was to use the n-gram data as an input for the Naive Bayes Classifier. The classifier used above was altered to process n-gram data instead of raw CFCs and run against the same dataset as above 620 samples of goodware and 94 samples of malware. In Figure 4 and 5 the results for 2-grams for goodware and malware are shown. Looking at these figures, we notice that the likelihoods of this classifier produces extremely low likelihoods for 2-gram. Most of the values are 0. We obtained the similar results with 4-gram and 6-gram experiments. Besides, from the statistical analysis in Section IV.A, we found that frequency is an important metric. It was observed that some n-grams appear more than once in a file and the frequency of occurrence was also recorded in the data files produced above. This occurrence data can now be used as a feature to abstract from the actual n-gram, which may differ because of compiler or encoder used. It was further reviewed if this feature can be used as a discriminator for identifying malware. The classifier was altered to filter this range of frequency and applied. Figures 6 and 7 show the testing results of the 10-50 frequency 2-grams.

| filename | prob. good | prob. bad | length of data |
|---|---|---|---|
| aspnet_compiler.lst | 5.37E-127 | 1.22E-130 | 266 |
| bootcfg.lst | 0 | 0 | 2913 |
| AdapterTroubleshooter.lst | 7.86E-026 | 2.50E-043 | 209 |
| BrmfRsmg.lst | 0 | 1.18E-246 | 1149 |

Fig. 4. Naïve Bayes classifier likelihoods for jcc goodware 2-gram

| filename | prob. good | prob. bad | length of data |
|---|---|---|---|
| bfb27f14234725a8f0146957953205ed.lst | 0 | 0 | 14373 |
| 1fd05f3185733f03e71543c0e27d7740.lst | 0 | 0 | 14877 |
| 0ec0f4be802b39a51c69bb0307a9629e.lst | 0 | 0 | 15187 |
| ff230a338ac820d73770411bab0013df.lst | 0 | 0 | 13106 |
| 6a8d6aec6af71a9ef65f4e1ac44da94b.lst | 0 | 0 | 14849 |
| 8f316e19714bad573af0f1116115cc33.lst | 0 | 0 | 14851 |
| 7e941465c1b5396697e9a2bebefe775c.lst | 0 | 0 | 15297 |
| 31f192e2e086723408ffc013bf546cbd.lst | 0 | 0 | 14875 |
| 9ecc6d7904710fd0b45926ae535a2529.lst | 0 | 0 | 14855 |
| 6fcc3a8b55376793f2985efbcb0123c8.lst | 0 | 0 | 14916 |
| 4c6a9aaaae5ec8cbb430a969bb17849c.lst | 0 | 0 | 13727 |
| 2fa2cbb2d273ab21aa1e10a6b314484f.lst | 0 | 0 | 14840 |
| 5a304d1f64643b9501f5d43a67460ca5.lst | 0 | 0 | 14853 |
| 425554e39f37bb5af1d8280e6fdd563d.lst | 0 | 0 | 13923 |

Fig. 5. Naïve Bayes classifier likelihoods for jcc malware 2-gram

| filename | prob. good | prob. bad | length of data |
|---|---|---|---|
| aspnet_compiler.lst | 0.5966469428 | 0.4033530572 | 0 |
| bootcfg.lst | 0.3239472531 | 1.1409411736 | 19 |
| AdapterTroubleshooter.lst | 0.5966469428 | 4.03E-001 | 0 |
| BrmfRsmg.lst | 0.1470200482 | 1.87E-001 | 2 |

Fig. 6. Naïve Bayes classifier likelihoods for jcc goodware 2-gram frequency

| filename | prob. good | prob. bad | length of data |
|---|---|---|---|
| bfb27f14234725a8f0146957953205ed.lst | 1.23E-011 | 6.36E-007 | 263 |
| 1fd05f3185733f03e71543c0e27d7740.lst | 1.05E-010 | 2.77E-006 | 289 |
| 0ec0f4be802b39a51c69bb0307a9629e.lst | 9.15E-011 | 1.89E-006 | 294 |
| ff230a338ac820d73770411bab0013df.lst | 2.04E-009 | 1.95E-005 | 207 |
| 6a8d6aec6af71a9ef65f4e1ac44da94b.lst | 1.05E-010 | 2.77E-006 | 288 |
| 8f316e19714bad573af0f1116115cc33.lst | 1.05E-010 | 2.77E-006 | 288 |
| 7e941465c1b5396697e9a2bebefe775c.lst | 1.05E-010 | 2.77E-006 | 300 |
| 31f192e2e086723408ffc013bf546cbd.lst | 1.05E-010 | 2.77E-006 | 289 |
| 9ecc6d7904710fd0b45926ae535a2529.lst | 3.81E-010 | 7.74E-006 | 288 |
| 6fcc3a8b55376793f2985efbcb0123c8.lst | 1.56E-010 | 3.40E-006 | 306 |
| 4c6a9aaaae5ec8cbb430a969bb17849c.lst | 1.41E-008 | 4.64E-005 | 260 |
| 2fa2cbb2d273ab21aa1e10a6b314484f.lst | 1.05E-010 | 2.77E-006 | 288 |
| 5a304d1f64643b9501f5d43a67460ca5.lst | 1.05E-010 | 2.77E-006 | 288 |
| 425554e39f37bb5af1d8280e6fdd563d.lst | 1.41E-008 | 5.54E-005 | 249 |

Fig. 7. Naïve Bayes classifier likelihoods for jcc malware 2-gram frequency

*C. Discussion*

From the experiments described above, there is no significant difference in the statistical values between these two groups using the median, variance or spread or derivatives of these values of control flow change data, they also show that there is a correlation between the statistical data of the CFC of goodware and malware except their frequency.

On the other hand, the experimental results show that data of the raw control flow change data used as training and test for a Naive Bayes Algorithm could not be used to distinguish malware from goodware. The reason is the goodware and malware samples were used as base "texts". Therefore the CFC features as "words" for training the algorithm and test samples were tested in the same manner, but the low result likelihoods showed that the feature pool was too big and the single significant features too scarce. Also, focusing on the single CFC feature, programs to be tested have to have enough single CFC features to provide a reliable basis for the comparison or likelihood calculation, as for example the Naive Bayes Algorithm will provide low values if very few features compared to the learned features are present in the to-test samples. More detailed analysis of the CFC features and a selection of features can improve the results.

Therefore, the use of more abstract features was tested, using the n-gram method to abstract from the raw CFC data. The n-grams were used to train the modified Naïve Bayes Algorithm and some data tested against it. Abstracting the data by extracting n-grams from the CFC data using 2-, 4- and 6- grams did not improve the results.

However, when looking at the frequency of the n-grams counted, it can improve the results from the raw CFC data. For example, the probability of goodware of AdapterTroubleshooter.lst (a goodware) is 0.59 vs. 0.004- its probability of malware. Looking at Figure 9, the probability of malware of all testing malwares is around 1000 times greater their probability of goodware.

## V. CONCLUSIONS AND FUTURE WORK

In this paper, we describe an approach of using the control flow change (CFC) itself and finding out if it is significant to classify 'goodware' and 'malware'. Statistical methods were used as well as the Naïve-Bayes algorithm to find out if there is signifi-cant data in CFC. It also was shown that data of the raw control flow change data used as training and test for a Naive Bayes Algorithm could be exploited to distin-guish malware from goodware. However, the data found showed at least the single feature chosen in some programs has been too scarce to be usable, and if data was found, the likelihoods were too low to make a decision. It is discovered that by far most of the CFC features are seen once to 5 times. That means that the single feature does have low significance. Therefore, in possible further approaches, it has to be combined with other features (for example all CFCs together) or grouped by similari-ty. The using of frequency of the n-grams can be used to distinguish malware from goodware with an appropriate threshold. However the overall likelihood is still low.

Therefore, in our further approach, using CFC could be to test if it can be used as a software family classifier or to detect code block or library reuse. Multiple confirmed samples of the same program family (e.g. malware families zeus or citadel) could be tested if the n-gram method reliably classifies these families together from test sets. We are also looking at advanced machine learning techniques such as [17][18][19] to compare with Naïve Bayes Classifier.